\documentclass{llncs}

\usepackage[dvips]{epsfig}

\begin{document}

\title{Evaluating the Impact of Information Distortion on Normalized
Compression Distance}

\author{Ana Granados, Manuel Cebri\'an, David Camacho
and Francisco B. Rodr\'iguez}

\institute{Departamento de Ingenier\'ia Inform\'atica, Universidad Auton\'oma de Madrid, Spain, \\
\email{(Ana.GranadosF,Manuel.Cebrian,David.Camacho,F.Rodriguez)@uam.es}}

\maketitle

\begin{abstract}

In this paper we apply different techniques of information
distortion on a set of classical books written in English. We
study the impact that these distortions have upon the Kolmogorov
complexity and the clustering by compression technique (the latter
based on Normalized Compression Distance, NCD). We show how to
decrease the complexity of the considered books introducing
several modifications in them. We measure how the information
contained in each book is maintained using a clustering error
measure. We find experimentally that the best way to keep the
clustering error is by means of modifications in the most frequent
words. We explain the details of these information distortions and
we compare with other kinds of modifications like random word
distortions and unfrequent word distortions. Finally, some
phenomenological explanations from the different empirical results
that have been carried out are presented.

\end{abstract}

\section{Introduction}
\label{Introduction}

A natural measure of similarity assumes that two objects $x$ and $y$
are similar if the basic blocks of $x$ are in $y$ and vice versa. If
this happens we can describe object $x$ by making reference to the
blocks belonging to $y$, thus the description of $x$ will be very
simple using the description of $y$.

This is what a compressor does to code the concatenated $xy$
sequence: a search for information shared by both sequences in
order to reduce the redundancy of the whole sequence.  If the
result is small, it means that a lot of information contained in
$x$ can be used to code $y$, following the similarity conditions
described in the previous paragraph. This was formalized by
Cilibrasi and Vit\'anyi \cite{Cilibrasi05}, giving rise to the
concept of \emph{Normalized Compression Distance} (NCD), which is
based on the use of compressors to provide a measure of the
similarity between the objects. This distance may then be used to
cluster those objects.

The mathematical formulation is as follows

\begin{equation}
NCD(x,y)=\frac{\max\{C(xy)-C(x),C(yx)-C(y)\}}{\max\{C(x),C(y)\}},
\end{equation}

where $C$ is a compression algorithm, $C(x)$ is the size of the
C-compressed version of $x$, and $C(xy)$ is the compressed size of
the concatenation of $x$ and $y$. NCD generates a non-negative
number $0 \leq NCD(x,y)\leq1$. Distances near 0 indicate
similarity between objects, while distances near 1 reveal
dissimilarity.

The theoretical foundations for this measure can be traced back to
the notion of Kolmogorov Complexity $K(X)$ of a string $X$, which
is the size of the shortest program able to output $X$ in a
universal Turing machine \cite{Turing36,Kolmogorov65,li1997ikc}.
As this function is incomputable due to the Halting problem
\cite{Sipser06}, the most usual estimation is based on data
compression: $C(X)$ is considered a good upper estimate of $K(x)$,
assuming that $C$ is a reasonably good compressor for $X$
\cite{Cilibrasi05}.

In our work we apply this distance to text clustering
\cite{Cilibrasi05,complearn}, with the aim to study the way in
which the method is influenced by different types of information
distortion. A percentage of words of the books is distorted by
using two different word-replacing techniques, which eventually
change the amount of information remaining in the books, their
(estimated) Kolmogorov Complexity, and the clustering error
obtained using the NCD.

Other authors \cite{Cebrian07} have given a theoretical and
experimental basis for explaining the NCD-clustering behavior of
elements which have been transmitted through a
\emph{symmetric-channel}, i.e. which have been perturbed by a
certain amount of uniform random noise. We go a step further by
considering a wider spectrum of information distortions, within
the framework of a complete experimental setup on a selected text
corpus for which an ideal clustering is already known.

The main contributions of this paper are \vspace*{-0.2cm}
\begin{itemize}
 \item New insights for the evaluation and explanation of the behavior of
the NCD-driven clustering method,
 \item a technique to reduce the Kolmogorov complexity of the
books while preserving most of the relevant information,
\item experimental evidence of how to fine-tune the NCD so that better clustering results
are obtained.
\end{itemize}
\vspace*{-0.2cm} This paper is structured as follows. Section
\ref{The Distortion Methods} describes the
distortion/word-replacement method, the clustering assessment and
the Kolmogorov Complexity estimation. Section \ref{Experiments and
Results} explains the experimental setup and gathers the results
of the experiments. Section \ref{Conclusions and Discussion}
summarizes the conclusions and describes ongoing research.

\section{The Distortion Methods}
\label{The Distortion Methods}

We want to study the effect of information distortion on
NCD-driven text clustering by replacing words from the documents
in different manners. After the distortion has been performed, we
execute the NCD clustering method on each distorted test set and
we quantitatively measure the error of the clustering results
obtained. Finally, the Kolmogorov complexity of the distorted
documents is estimated, 
based on the concept that data compression is an upper bound of
the Kolmogorov complexity.

\subsection{Replacement methods}
\label{Replacement methods}

We use six different replacement methods, which are pairwise
combinations of two factors: \emph{word selection} and
\emph{substitution method}. \vspace*{-0.2cm}
\begin{itemize}
 \item Word selection: we incrementally select a percentage $p$, and we eliminate the
       $p$-\emph{most}/\emph{least}/\emph{randomly} frequent words in English from the books. We estimate the frequencies
       of words in English using the British National Corpus \cite{BNC}.
\item Substitution method: each character of the word that is distorted according to the word-frequency, is
      substituted by either a \emph{random character} or an \emph{asterisk}.
\end{itemize}
\vspace*{-0.2cm} Note that all six combinations maintain the
length of the document. This is enforced to ease the comparison of
the Kolmogorov Complexity among several methods.

\subsection{Assessing the Clustering}
\label{Assessing the Clustering}


The CompLearn Toolkit \cite{complearn} implements the clustering
method described in \cite{Cilibrasi05}. The clustering method
comprises two phases. First, the NCD matrix is calculated using
the LZMAX compressor, a Lempel-Ziv-Markov chain Algorithm
\cite{LZMA}. Second, the NCD matrix is used as input to the
clustering phase and a dendrogram is generated as an output. A
dendrogram is an undirected binary tree diagram, frequently used
for hierarchical clustering, that illustrates the arrangement of
the clusters produced by a clustering algorithm.

\begin{figure}[htb]
  \includegraphics[width=8.3cm]{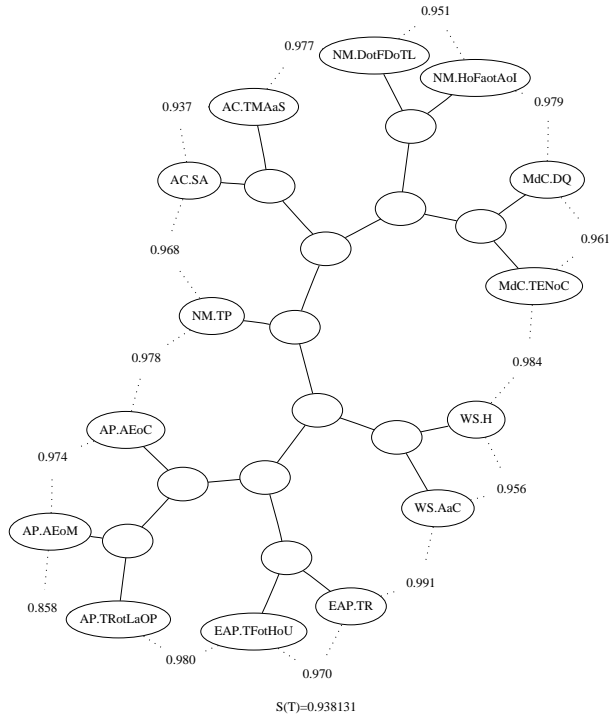}
  \centering
  \caption{Example of dendrogram.}\label{Fig:dendro}
\vspace*{-0.5cm}
\end{figure}

In Fig \ref{Fig:dendro} we can observe one of the dendrograms that
we have obtained. Each leaf of the dendrogram corresponds to a
document, and has a label that starts with the acronym of the
author, and ends with the acronym of the title. For example, the
node with label $AP.AEoM$ corresponds to the document \emph{An
Essay on Man} by Alexander Pope.

Once the CompLearn Toolkit is used to cluster the
documents, we need to quantitatively evaluate the error of the
dendrograms obtained. We define the distance between two nodes as
the minimum number of internal nodes needed to go from one to the
other. For example, in Fig \ref{Fig:dendro} the distance between
the nodes with label $WS.H$ and $WS.AaC$ would be one, since both
nodes are connected to the same internal node. We use this concept
to measure the error of a dendrogram.


We add all the pairwise distances between nodes starting with the
same string, i.e. we add all the pairwise distances between the
documents by the same author. For example, in Fig \ref{Fig:dendro}
there are three nodes which label starts with $AP$, thus we add
the distance between $AP.AEoC$ and $AP.AEoM$, between $AP.AEoC$
and $AP.TRotLaOP$, and between $AP.AEoM$ and $AP.TRotLaOP$. We
repeat this procedure with the nodes of each author, obtaining a
certain total quantity. The bigger the measure, the worse the
clustering.

The ideal dendrogram would be a clustering where all the documents
by the same author are grouped together. The clustering error
corresponding to the ideal dendrogram is 14 for these documents.
Note that in Fig \ref{Fig:dendro} the node with label \emph{NM.TP}
is clustered incorrectly. Thus, the clustering error corresponding
to this dendrogram is 18 instead of 14.

\section{Experiments and Results}
\label{Experiments and Results}

The experiments have been designed to evaluate the impact of
information distortion on NCD-driven text clustering by
incrementally replacing words from the documents in different
manners. We measure the error of the clustering in presence of
distortion and compare it with two baselines: the ideal
clustering, and the non-distorted NCD-driven clustering.

We have applied the NCD clustering method over a set of fourteen
classical books written in English. We have two books by Agatha
Christie: \emph{The Secret Adversary}, and \emph{The Mysterious
Affair at Styles}. Three books by Alexander Pope: \emph{An Essay
on Criticism}, \emph{An Essay on Man}, and \emph{The Rape of the
Lock, an heroic-comical Poem}. Two books by Edgar Allan Poe:
\emph{The Fall of the House of Usher}, and \emph{The Raven}. Two
books by Miguel de Cervantes: \emph{Don Quixote}, and \emph{The
Exemplary Novels}. Three books by Niccol\`o Machiavelli:
\emph{Discourses on the First Decade of Titus Livius},
\emph{History of Florence and of the Affairs of Italy}, and
\emph{The Prince}. Two books by William Shakespeare: \emph{The
tragedy of Antony and Cleopatra}, and \emph{Hamlet}.



We show the results of clustering the classical books when the six
different replacement methods are used to preprocess them. Every
figure plots the clustering error or the complexity of the books
vs. a certain percentage of replaced words. The curve with
asterisk markers represents the results obtained when the
characters of the words are replaced with asterisks. The curve
with square markers corresponds to the results obtained when
random characters are used to replace the characters of the
distorted words. Furthermore, two constant lines may appear in
each figure. One corresponds to the measure in the ideal
clustering and the other corresponds to the non-distorted
NCD-driven clustering. The former is 14, the latter is 18.

In every graph, the value on the horizontal axis corresponds to
the fraction of the total BNC frequency that is associated to the
words being distorted. Note that even when all the words included
in the BNC are replaced from the texts, the words that are not
included in the BNC remain in the documents. For example, in the
book \emph{Don Quixote} by Miguel de Cervantes, words like
\emph{Dulcinea} or \emph{Micomicona}, the names of two characters,
remain in the documents when all the words of the BNC are
distorted from the documents.

\begin{figure}[h]
  \rotatebox{270}{\resizebox{!}{8cm}{%
   \includegraphics{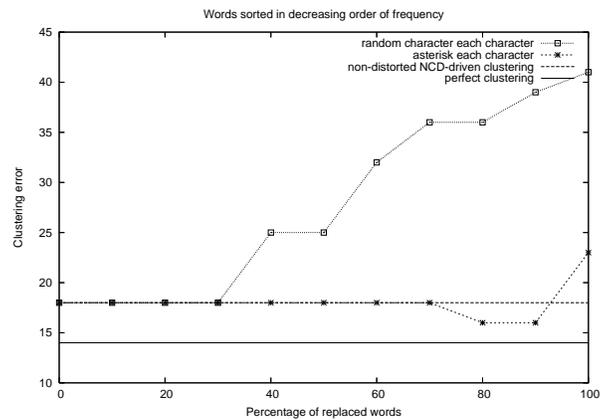}}}
  \centering
  \caption{Clustering error. Words sorted in decreasing order of frequency.}\label{Fig:ord-mantienen}
\vspace*{-0.35cm}
\end{figure}

\begin{figure}[h]
  \rotatebox{270}{\resizebox{!}{8cm}{%
   \includegraphics{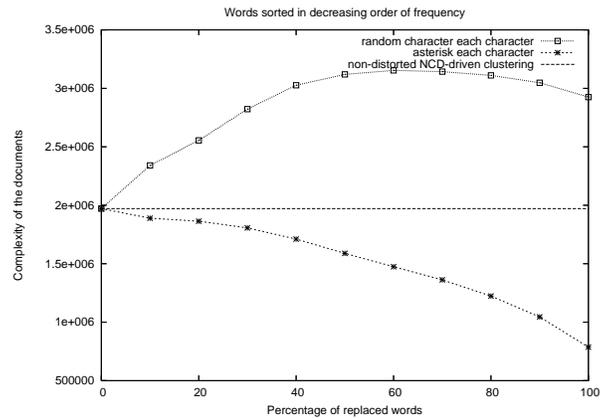}}}
  \centering
  \caption{Complexity of the documents. Words sorted in decreasing order of frequency.}\label{Fig:complex-ord-mantienen}
\vspace*{-0.5cm}
\end{figure}

\begin{figure}[h]
\rotatebox{270}{\resizebox{!}{8cm}{%
   \includegraphics{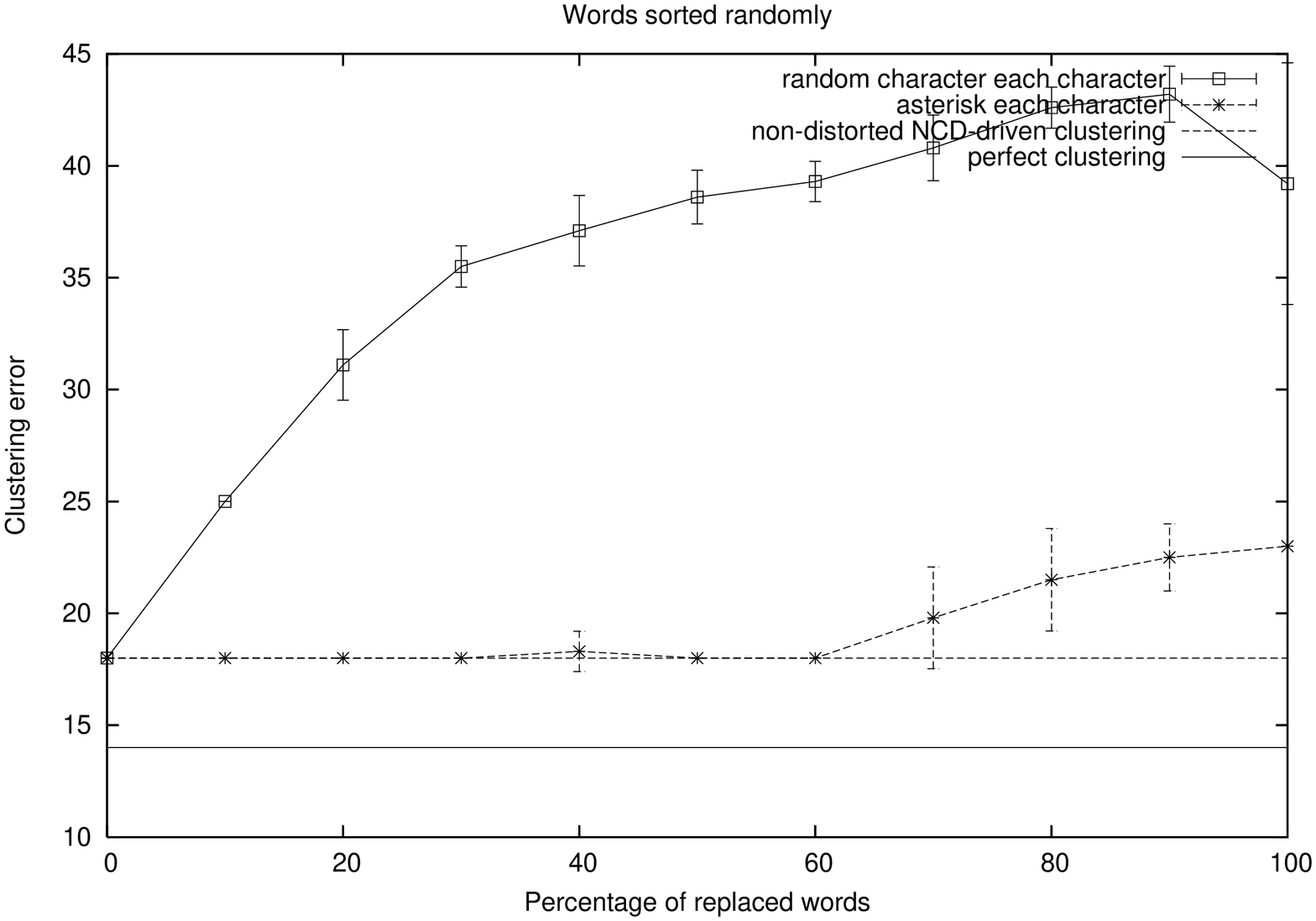}}}
  \centering
  \caption{Clustering error. Words sorted randomly (mean and standard deviation.)}\label{Fig:desord-mantienen}
\vspace*{-0.35cm}
\end{figure}

\begin{figure}[h]
\rotatebox{270}{\resizebox{!}{8cm}{%
   \includegraphics{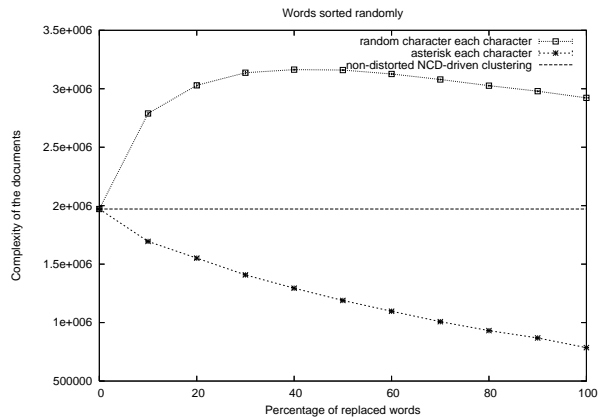}}}
  \centering
  \caption{Complexity of the documents. Words sorted randomly (mean and standard deviation.)}\label{Fig:complex-desord-mantienen}
\vspace*{-0.5cm}
\end{figure}

\begin{figure}[h]
\rotatebox{270}{\resizebox{!}{8cm}{%
   \includegraphics{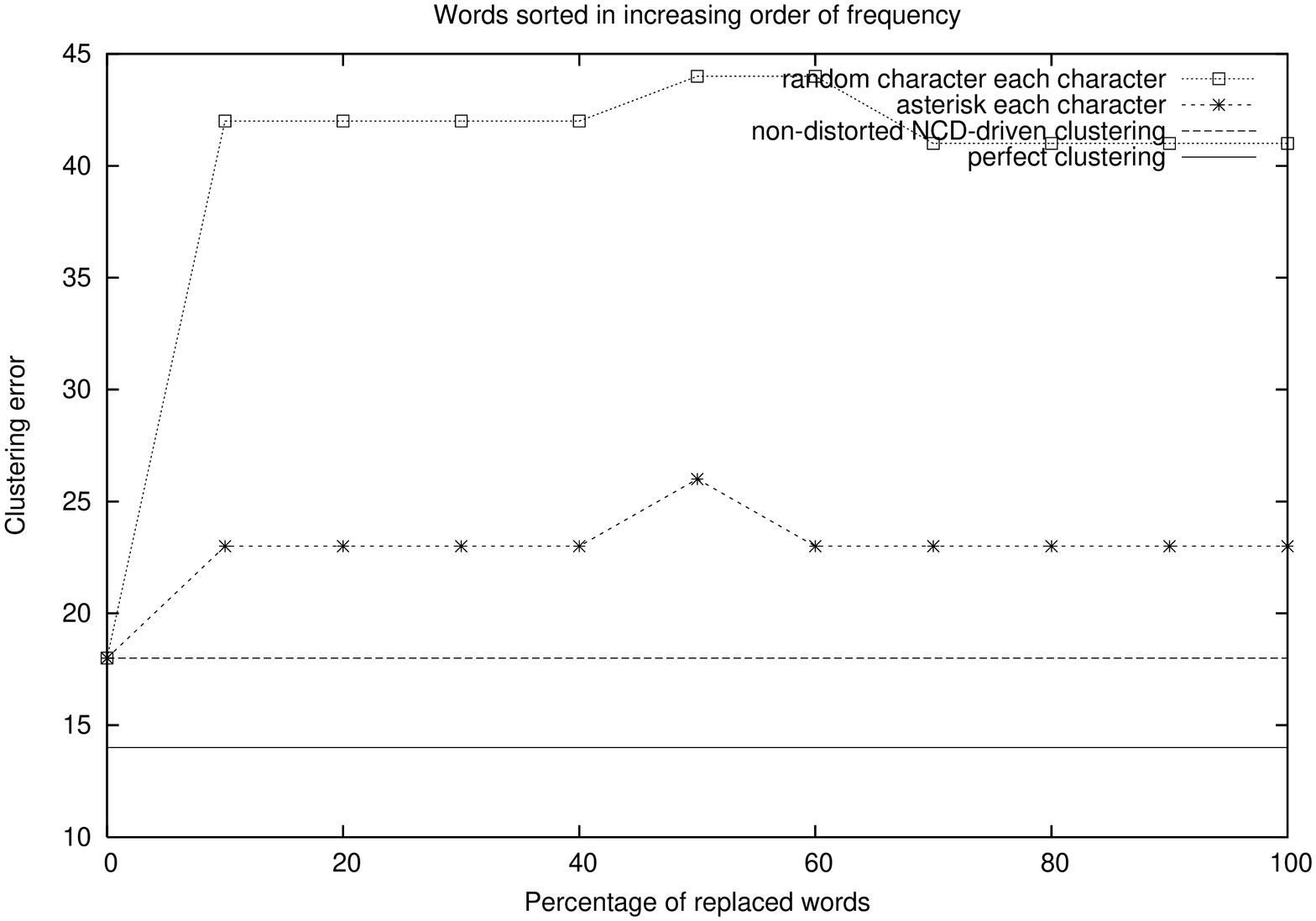}}}
  \centering
  \caption{Clustering error. Words sorted in increasing order of frequency.}\label{Fig:reves-mantienen}
\vspace*{-0.35cm}
\end{figure}

\begin{figure}[h]
\rotatebox{270}{\resizebox{!}{8cm}{%
   \includegraphics{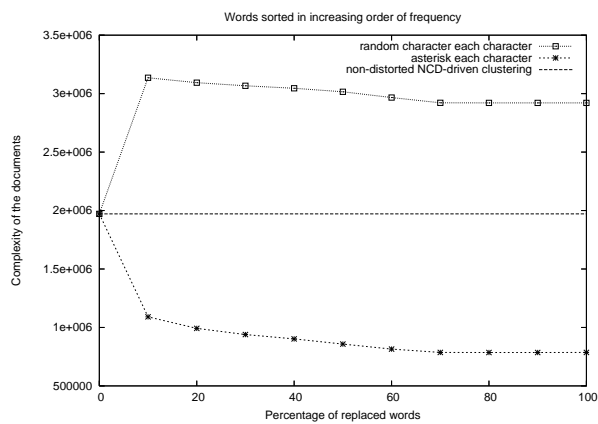}}}
  \centering
  \caption{Complexity of the documents. Words sorted in increasing order of frequency.}\label{Fig:complex-reves-mantienen}
\vspace*{-0.5cm}
\end{figure}

The clustering error vs. the percentage of replaced words is presented
in Figs \ref{Fig:ord-mantienen}, \ref{Fig:desord-mantienen} and \ref{Fig:reves-mantienen},
which show the results for the $X\%$-\emph{most}/\emph{randomly}/\emph{least} frequent words respectively. Figs
\ref{Fig:complex-ord-mantienen},
\ref{Fig:complex-desord-mantienen} and
\ref{Fig:complex-reves-mantienen} show the evolution of the
complexity of the documents as a function of the same percentages.

In Fig \ref{Fig:ord-mantienen} we observe that when the characters
of the words are replaced with random characters the clustering
error increases. When the characters are replaced with asterisks
the clustering error remains stable. If we observe the curve with
asterisk markers at 80\% and 90\%, we can see that the results are
better than those obtained for the non-distorted documents,
although they are not as good as those that would correspond to
the ideal clustering.

Looking at Fig \ref{Fig:complex-ord-mantienen}, we realize that
the complexity of the documents rises when the substitution method
is based on random characters. However, when it is based on
asterisks the complexity of the documents decreases, because a
great amount of characters from the documents are replaced with
the same character, which increases the redundancy of the document
and thus makes it more compressible.

Fig \ref{Fig:desord-mantienen} shows the mean and the standard
deviation of the results obtained in ten different experiments.
The clustering error increases when random character substitution
is applied. However, when asterisk substitution is applied the
error keeps stable until 60\%. From 60\% to 100\% the error
increases. Comparing Figs \ref{Fig:ord-mantienen} and
\ref{Fig:desord-mantienen} we observe that better results are
obtained when we start disturbing the most frequent words. This
makes us think that the frequency of the replaced words could
affect the clustering.

The mean and the standard deviation of the complexity are
presented in Fig \ref{Fig:complex-desord-mantienen}, although the
standard deviation is difficult to visualize due to the fact that
its absolute value is very small as compared to the mean. Note
that this graph decreases faster than the graph which represents
the complexity when we start replacing the most frequent words,
see Fig \ref{Fig:complex-ord-mantienen}.

When the characters of the words are replaced with random
characters, as shown in Fig \ref{Fig:reves-mantienen}, the
clustering error increases faster than before, see Figs
\ref{Fig:ord-mantienen} and \ref{Fig:desord-mantienen}. When the
words are replaced with asterisks the clustering error increases
rapidly and then remains stable. This phenomenon could be due to
the fact that when we start replacing the least frequent words, we
replace precisely those words which carry the most information in
terms of clustering compression.

When the substitution method is based on random characters, as
shown in Fig \ref{Fig:complex-reves-mantienen}, the complexity of
the documents grows sharply compared to the evolution observed in
Figs \ref{Fig:complex-ord-mantienen} and
\ref{Fig:complex-desord-mantienen}. When the substitution method
is based on asterisks the complexity of the documents decreases
sharply as compared to the same figures. This is due to the fact
that when we start disturbing the X\%-least frequent words, lots
of words are required to achieve the 10\% of the BNC frequencies.

\begin{figure}[h]
\rotatebox{270}{\resizebox{!}{8cm}{%
   \includegraphics{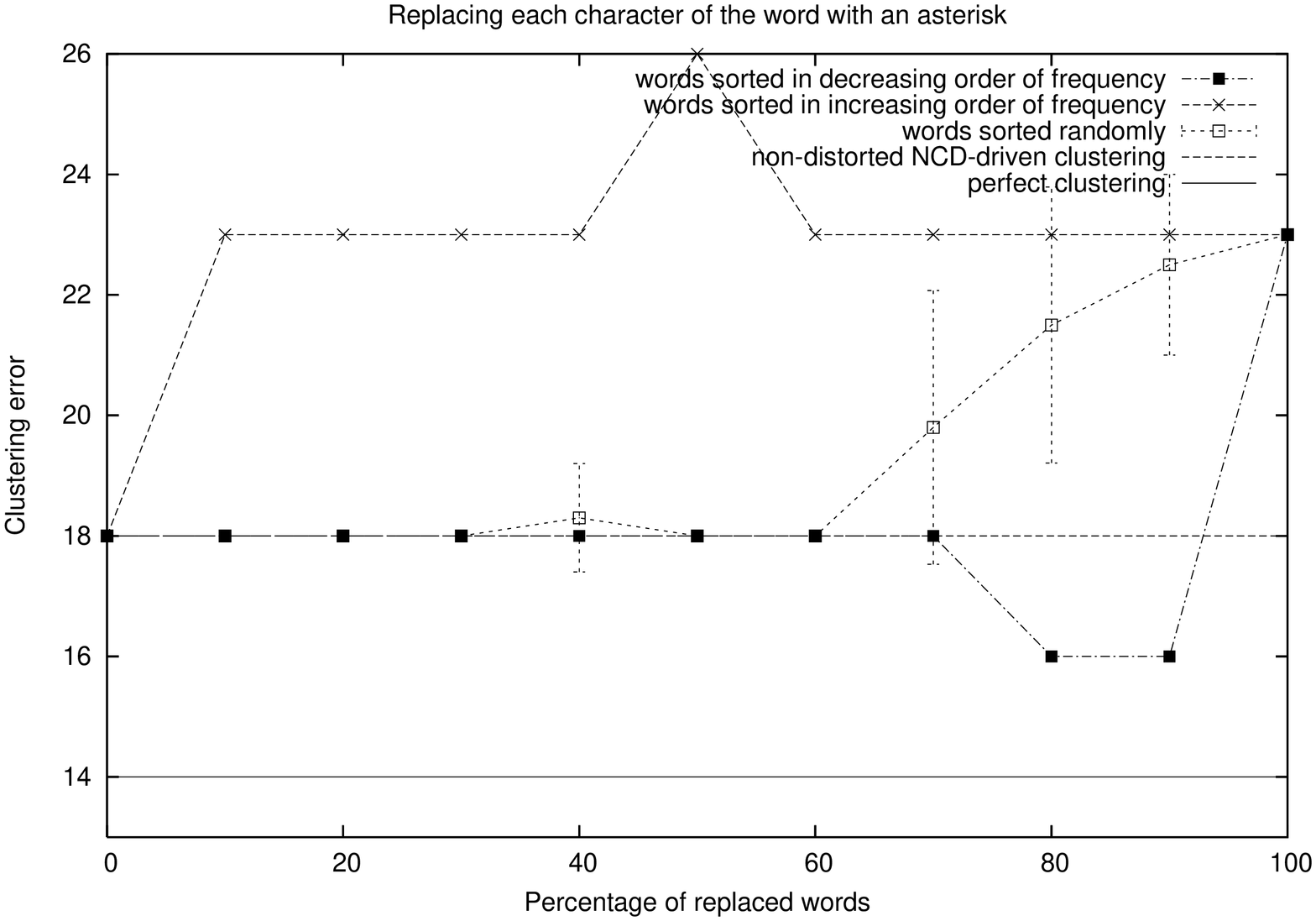}}}
  \centering
  \caption{Clustering Results. Asterisk each character of the replaced word.}\label{Fig:asteriscos}
\vspace*{-0.35cm}
\end{figure}

In order to give a better comparison we illustrate in Fig
\ref{Fig:asteriscos} the clustering error obtained when the words
are replaced with asterisks for the three different word
selections: $p$-\emph{most}/\emph{least}/\emph{randomly} frequent
words in English. We observe that the better results are obtained
when we start distorting the $p$-\emph{most} frequent ones, and
the worst results are obtained then we start distorting the
$p$-\emph{least} frequent ones. When we select randomly the words
the results keep between the others. These facts empirically
demonstrate that the frequency of the words affects the clustering
results when we cluster these books using the CompLearn Tool with
the LZMAX compressor.

\begin{figure}[h]
\rotatebox{270}{\resizebox{!}{8cm}{%
   \includegraphics{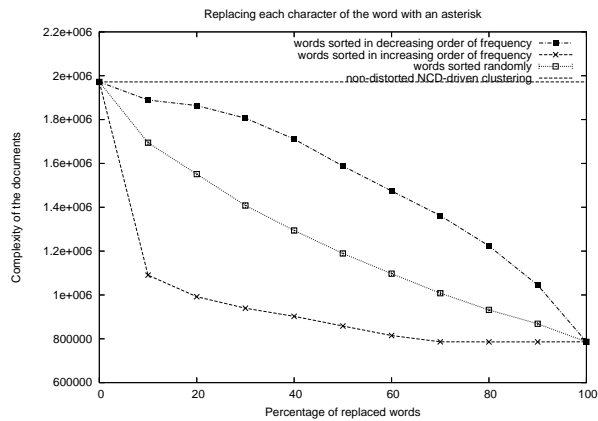}}}
  \centering
  \caption{Complexity of the documents. Asterisk each character of the replaced word.}\label{Fig:complex-asteriscos}
\vspace*{-0.5cm}
\end{figure}

In an analogous way, the complexity of the documents for the three
word selection techniques is depicted in Fig
\ref{Fig:complex-asteriscos}. Comparing Figs \ref{Fig:asteriscos}
and \ref{Fig:complex-asteriscos} we observe that document
complexity and clustering error are negatively correlated. Thus,
the best clustering results are obtained using the word selection
that reduces the least the complexity of the documents.

To summarize, when random characters are used to replace the words
in the text preprocessing phase, the error of the clustering
method increases with the percentage of words removed,
independently of the word selection used. When the words are
replaced using asterisks the clustering error is always smaller
than the one obtained when using random characters. Furthermore,
the best clustering results are obtained when we select the most
frequent words, and the substitution method is based on asterisks.
In this case, for 80\% and 90\%, the results obtained from the
original texts are improved. Moreover, comparing all the figures,
it can be observed that the frequency of the removed words has an
influence over the clustering error.


\section{Conclusions and Discussion}
\label{Conclusions and Discussion}

We have applied the clustering method detailed in
\cite{Cilibrasi05} to cluster several English classical books. We
have studied how the clustering method is affected by different
types of information distortion. In order to do that, we have
measured the clustering error vs. the percentage of words
distorted. The Kolmogorov complexity of the books has been
estimated as well, to study the impact of the information
distortion on the complexity of the documents.

Although several distortion methods have been designed, in this
paper we have only considered those which maintain the initial
length of the books to ease the comparison of the Kolmogorov
complexity among them.

Three main contributions have been presented in this paper. First,
we have made an empirical evaluation of the behavior of the
NCD-driven clustering method, and the way in which an incremental
modification of the books affects the clustering error. Second, we
have presented a technique which reduces the Kolmogorov complexity
of the books while preserving the relevant information therein.
Third, we have observed experimental evidence of how to improve
the NCD-clustering method by preprocessing the books in a certain
manner.

The experimental results show how the clustering error is
maintained even when the information contained in the documents is
reduced progressively by replacing the words using a special
character. We have found that replacing the most frequent words
gives us the better clustering results. This method maintains the
clustering error with very high values of distortion, and even
improves the non-distorted NCD-driven clustering when the
80\%-90\% of the words are replaced from the documents, see Fig
\ref{Fig:ord-mantienen}. This means that we are replacing exactly
non-relevant parts of the books. This makes it easier for the
compressor to estimate the complexity of the documents in an
accurately manner. Therefore, the compressor obtains more reliable
similarities.

Other techniques, such as randomly selecting the words to replace,
or replacing the least frequent ones have been studied and
analyzed. Despite the complexity of the documents being reduced
too (see Figs~\ref{Fig:complex-desord-mantienen},
and~\ref{Fig:complex-reves-mantienen}), the clustering error
increases faster (see Figs~\ref{Fig:desord-mantienen},
and~\ref{Fig:reves-mantienen}). Thus, the information that has
been replaced is relevant in the clustering process, and
consequently we are losing important information. Therefore, the
similarities among the documents are not being correctly measured.

In the future, we plan to work in several ways to study the
observed behavior in other textual repositories, like scientific
documentation, or genome-based repositories. However, the
NCD-based clustering is a general technique so it is possible to
use other kinds of sources, such as, music or video. In these
domains it would be necessary to analyze how the distortion method
could be designed. Other well-known compression algorithms, like
PPMZ, BZIP2 or GZIP, will be analyzed to evaluate if the
complexity estimation affects the clustering behavior as much as
it does in other NCD-driven experiments \cite{Cebrian05}. Finally,
we will apply these techniques in other areas like Information
Retrieval.

\section*{Acknowledgment}
This work was supported by TIN 2004-04363-CO03-03, TIN 2007-65989,
CAM S-SEM-0255-2006, TIN2007-64718 and TSI 2005-08255-C07-06. We
would also like to thank Franscico S\'anchez for his useful
comments on this draft.

\bibliographystyle{splncs}
\bibliography{vitacora}

\begin{thebibliography}{10}

{\scriptsize

\bibitem{Cilibrasi05}
Cilibrasi, R., Vitanyi, P.:
\newblock Clustering by compression.
\newblock IEEE Transactions on Information Theory \textbf{51}(4) (2005)
  1523--1545

\bibitem{Turing36}
Turing, A.:
\newblock On computable numbers, with an application to the
  entscheidungsproblem.
\newblock Proceedings of the London Mathematical Society \textbf{2}(42) (1936)
  230--265

\bibitem{Kolmogorov65}
Kolmogorov, A.:
\newblock Three approaches to the quantitative definition of information.
\newblock Problems Information Transmission \textbf{1}(1) (1965)  1--7

\bibitem{li1997ikc}
Li, M., Vit{\'a}nyi, P.:
\newblock {An introduction to Kolmogorov complexity and its applications}.
\newblock Springer-Verlag Graduate Texts In Computer Science Series (1997)  637

\bibitem{Sipser06}
Sipser, M.:
\newblock Introduction to the Theory of Computation. Second edn.
\newblock PWS Publishing (2006)

\bibitem{complearn}
Cilibrasi, R., Cruz, A.L., de~Rooij, S., Keijzer, M.:
\newblock CompLearn Toolkit.
\newblock [Online] Available: http://www.complearn.org/

\bibitem{Cebrian07}
Cebri\'an, M., Alfonseca, M., Ortega, A.:
\newblock The normalized compression distance is resistant to noise.
\newblock IEEE Transactions on Information Theory \textbf{53}(5) (2007)
  1895--1900

\bibitem{BNC}
Consortium, B.N.C.:
\newblock British National Corpus.
\newblock Oxford University Computing Services [Online] Available:
  http://www.natcorp.ox.ac.uk/

\bibitem{LZMA}
Pavlov, I.:
\newblock LZMA.
\newblock [Online] Available: http://www.7-zip.org/sdk.html

\bibitem{Cebrian05}
Cebri\'an, M., Alfonseca, M., Ortega, A.:
\newblock Common pitfalls using normalized compression distance: what to watch
  out for in a compressor.
\newblock Communications in Information and Systems \textbf{5}(4) (2005)
  367--384
}

\end{thebibliography}
\end{document}